# Integrated lithium niobate optical phased array for two-dimensional beam steering

**GONGCHENG YUE, YANG LI***

*State Key Laboratory of Precision Measurement Technology and Instruments, Department of Precision Instrument, Tsinghua University, Beijing, 100084 China*
*\*Corresponding author: yli9003@mail.tsinghua.edu.cn*

**Optical phased arrays (OPAs) with high speed, low power consumption, and low insertion loss are appealing for many applications, such as light detection and ranging, free-space communication, image projection, and imaging. These OPAs can be achieved by fully leveraging the high electro-optic modulation speed, low driving voltage, and low optical loss of integrated lithium niobate (LN) photonics. Here we present an integrated LN OPA operating in the near-infrared regime, demonstrating a 24×8° two-dimensional beam steering, a far-field beam spot with full-width at half-maximum of 2×0.6°, and a side lobe suppression level of 10 dB. Additionally, our OPA's phase modulator features a half wave voltage of 6 V. Our OPA opens the door of high speed, low power-consumption, and low loss integrated LN OPAs for various applications.**

Optical phased arrays (OPAs) have broad applications in light detection and ranging, free-space communication, image projection and imaging[1-4]. They have been implemented based on various mechanisms, including liquid crystal, microelectromechanical systems (MEMS), and integrated photonics[5-8]. So far, integrated photonics-based OPAs have been realized in several material platforms including silicon, silicon nitride, and indium phosphide[9-11]. Because silicon offers the unique advantage of mass production via CMOS compatible fabrication process[12, 13], so far most integrated photonics-based OPAs are fabricated on silicon platform[14-16] with the phase of each channel tuned by thermo-optic effect. However, this tuning method has limited modulation speed, typically in the megahertz range, and high-power consumption especially when the number of phase modulators reaches hundreds to thousands. To overcome these limitations, silicon electro-optic modulators have been proposed as an alternative[10, 17], offering higher modulation speed and lower power consumption. However, this approach introduced a higher insertion loss from the larger propagation loss of the p-n junction waveguide.

Thin-film lithium niobate (LN) is an emerging material platform for simultaneously achieving high modulation speed, low power consumption and low optical propagation loss[18, 19]. Owing to the high index contrast of etched LN waveguide, integrated LN modulator can achieve a narrow electrode-waveguide spacing, leading to ultra-low driving voltage[20]. Furthermore, integrated LN modulators demonstrated modulation speeds exceeding 100 GHz[21, 22], showing a great potential in realizing high-speed OPAs. Moreover, LN waveguide's optical propagation loss can be as low as 0.027 dB/cm[23], showing a significant advantage over silicon and indium phosphide. Additionally, LN exhibits a wide transparent window spanning from 350 nm to 5000 nm[24], enabling OPAs operating from visible to mid-infrared spectra.

Here, we present a 16-channel integrated LN OPA operating in near-infrared regime. Our OPA utilizes electro-optic modulation and wavelength tuning to achieve beam steering over 24° and 8°, respectively, in two dimensions. Our experimental results show that the side lobe suppression level is below 10 dB while the beam is steering from −12° to +12°. Moreover, the FWHM of far-field beam spot is 2×0.6°. To the best of our knowledge, we present the first research article on integrated LN OPA. Previous to our work, there is one conference proceeding briefly reporting an integrated LN OPA based on electro-optic modulation[25]. Beyond this work, we provide a systematic design analysis of LN-based OPA. Moreover, we study the influence of photorefractive effect in LN-based OPA and provide several potential solutions to mitigate this effect.

Our OPA consists of waveguide for guiding light, multimode interferometers (MMI) for optical power division, electro-optic modulators for phase modulation, and grating waveguides for out-of-plane radiation (Fig. 1a). This OPA is based on a 300-nm-thick X-cut thin film LN wafer. Light is first coupled into LN waveguide through edge coupling, and is subsequently split into 16 channels by cascaded MMI. The phase of light in each channel is modulated by the electric field which is exerted by the aluminum electrodes. Such electrodes are connected to bonding pads through vias. The bonding pads are wire-bonded to a printed circuit board (PCB) that is connected to a voltage source (Fig. 1b). After phase modulation, all the channels are guided to closely spaced grating waveguide array, leading to the out-of-plane radiation. The steering directions controlled by phase modulation and wavelength tuning are denoted as $\theta_x$ and $\theta_y$, respectively (Fig. 1a).

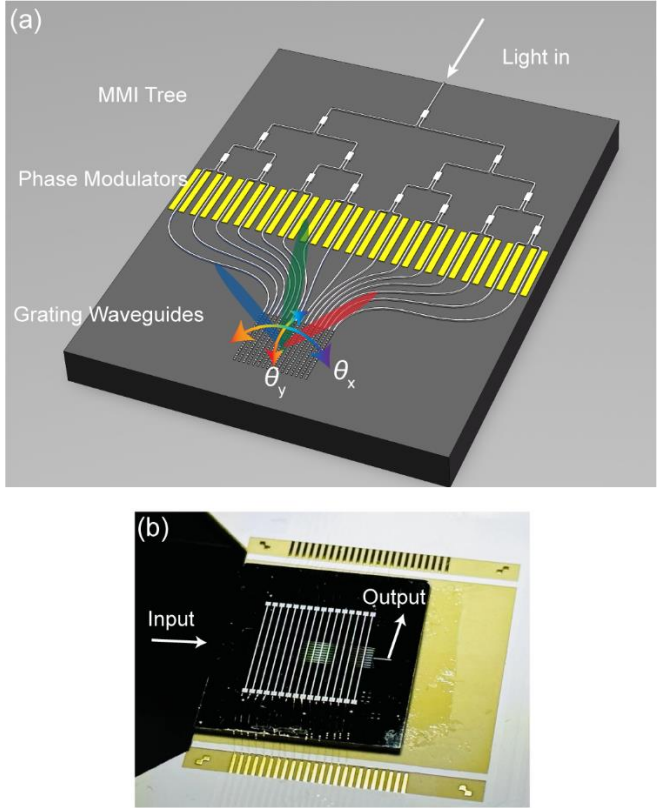

Fig. 1. Integrated LN OPA. (a) Schematic of the device. (b) Optical image of the OPA chip wire-bonded to a PCB.

Our OPA uses cascaded MMIs to equally split light to 16 channels (Fig. 2a). For each MMI, the 1-μm-wide strip waveguide is tapered up to a width of 4 μm before coupling into the multimode interference region, reducing the insertion loss of MMI. Based on the analytical theory[26], we designed the width and length of the multimode interference region as 10 μm and 63 μm, respectively. Finite difference time domain (FDTD) simulation results show that this MMI features an average insertion loss lower than 0.1 dB in the wavelength regime from 1500 nm to 1600 nm. Based this design, we fabricated these MMIs (Fig. 2b). After splitting light to 16 channels, light of each channel is modulated by an electro-optic phase modulator to realize beam steering along $\theta_x$ direction (Fig. 1a).

Each electro-optic phase modulator is composed of 10-mm long electrodes along a strip waveguide. As shown in the top left inset of Fig. 2a, aluminum electrode with a 150-nm thickness is first lifted off with a 3-μm spacing to the waveguide. 1.5-μm silicon dioxide is then deposit above the waveguides and electrodes using plasma enhanced chemical vapor deposition. Next, we patter resist above the silicon dioxide and then etch the silicon dioxide to open windows at vias' locations. After opening windows, second layer of the electrodes is lifted off. This layer of electrodes connects the bottom layer of electrodes and the bonding pads through vias (Top left inset of Fig. 2a, Fig.2 c). For each phase modulator, the simulated product of half wave voltage and length is 6 V*cm. This value is higher than that of the state-of-the-art LN phase modulators consisting of ridge waveguide[27] because the modulation strength of strip waveguide-based modulator is lower than that of the ridge waveguide-based modulator[28]. The product of half wave voltage and length can be further reduced to 2 V*cm when the phase modulator operating at visible wavelengths[29]. The presence of the electrodes leads to a 100-μm minimum spacing between adjacent waveguides of phase modulators, necessitating the routing of waveguides to the grating-waveguide region where the waveguides are very close to each other.

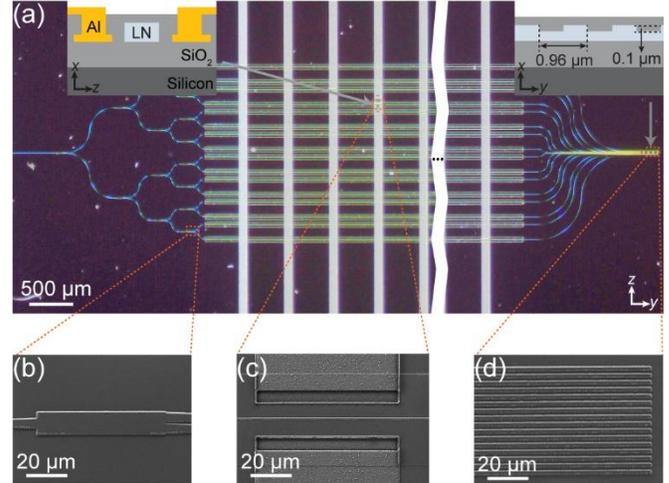

Fig. 2. Optical microscope image and SEM images of our OPA chip. (a) Optical microscope image of our OPA chip. Top left inset shows the cross-sectional view in the *x-z* plane of the phase modulator. Top right inset shows the cross-sectional view in the *x-y* plane of the grating waveguide. (b) SEM image of the MMI. (c) SEM image of the modulator in the vias region. (d) SEM image of the grating waveguides.

To increase the field of view (FOV) of the phase tuning-induced beam steering, a smaller spacing between grating waveguides is desirable. However, if grating waveguides are too close to each other, there will be strong crosstalk between waveguides. Therefore, we choose strip waveguide instead of ridge waveguide as the grating waveguide because strip waveguide can provide a better mode confinement, leading to a lower crosstalk between closely spaced waveguides. Considering the fact that a wider LN waveguide gives a lower propagation loss, we designed the grating waveguide based on 1-μm-wide strip waveguide to achieve a good balance among propagation loss, FOV, and crosstalk. To simplify the fabrication process, we also chosen the 1-μm-wide strip waveguide as the design of guiding waveguides in the MMI and modulation regions. LN OPA usually needs a larger spacing between grating waveguides than that of silicon OPA because of the weaker mode confinement of LN waveguide.

We designed the grating waveguide in the form of double etched grating waveguide (Top right inset of Fig. 2a). In general, there are two kinds of grating waveguide for OPA — corrugated grating waveguide and double etched grating waveguide[30]. Corrugated grating waveguide achieves radiation by changing width, which can simplify the fabrication process through one-time patterning and etching. Double etched grating waveguide modulates height to realize radiation. It requires a more complicated fabrication process including two times of patterning and etching. However, the crosstalk between corrugated grating waveguides is higher than that of the double etched grating waveguides[31], necessitating a larger spacing between grating waveguides. To reduce the spacing between grating waveguides, we chosen double etched waveguide. We designed such a waveguide with 0.96-μm period, 0.5 duty cycle, 0.1-μm double etching depth, and 3-μm lateral center-to-center spacing (Top right inset of Fig. 2a, Fig. 2c). FDTD simulation results show that our design features a crosstalk between adjacent grating waveguides lower than 2.5% (Fig. 3b).

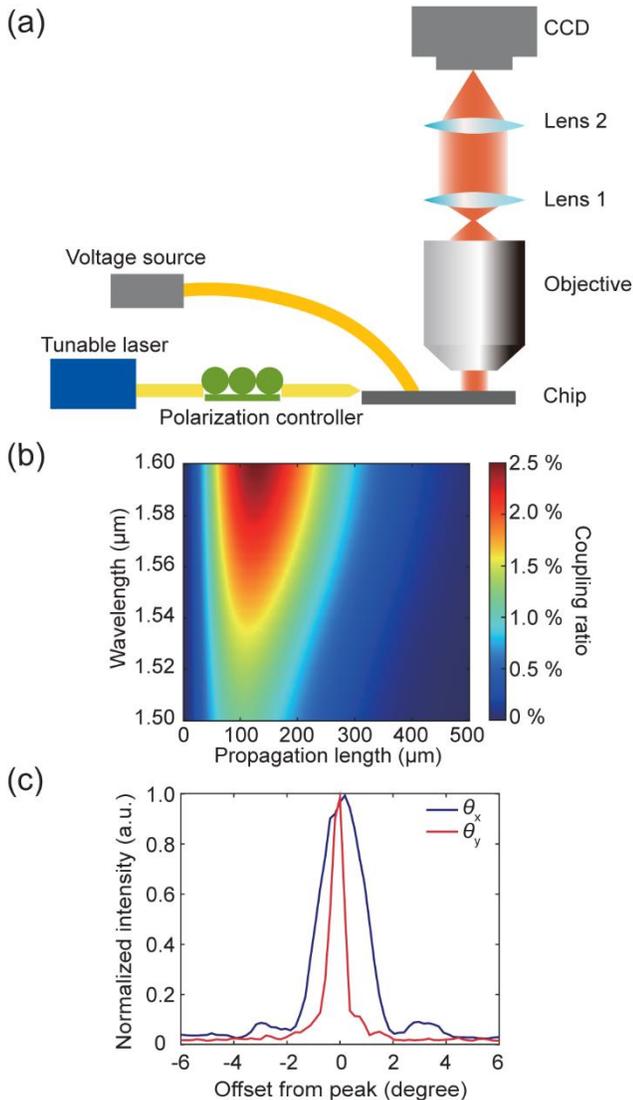

Fig. 3. Experiment setup and performance of grating waveguides. (a) Experiment setup for far-field measurement. (b) Simulated coupling ratio from one grating waveguide to the adjacent grating waveguide. (c) Cross sections of measured far-field beam spot. The FWHMs of cross sections over $\theta_x$ and $\theta_y$ directions are 2° and 0.6°, respectively.

To characterize the far-field radiation of our OPA, we built a conventional Fourier imaging system (Fig. 3a)[32]. The microscope objective (NA =0.4) performs the Fourier transform from near field to far field in the Fourier plane. Then, two lenses image the Fourier plane to the infrared CCD. Because of the random distribution of the initial phases of all the channels, the initial far-field pattern of radiation is randomly distributed. To condense the beam spot to the desired position in far field (Fig. 3c), we used a hill-climber-based optimization algorithm to search for the optimal voltage for each channel. This beam spot's FWHMs over cross sections along $\theta_x$ and $\theta_y$ directions are 2° and 0.6°, respectively. There is a tradeoff between FWHM over $\theta_x$ direction and FOV. I.e. A larger lateral spacing between grating waveguides will decrease the FWHM in $\theta_x$ direction but will decrease the FOV.

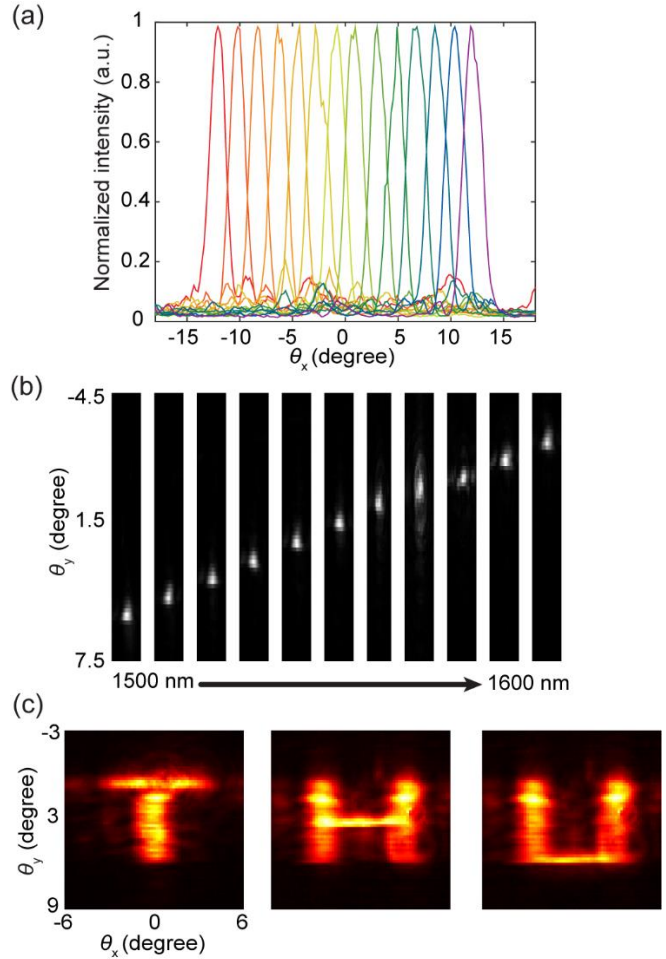

Fig. 4. Measured beam-steering results. (a) Measured far-field beam patterns over $\theta_x$ cross section. The steering range is −12° to +12° in which each measurement is depicted using a distinct color. We steer the beam to a desired angle via electro-optic phase modulation. (b) Images of far-field beam patterns. We steer the beam along $\theta_y$ direction by tuning the wavelength from 1500 nm to 1600 nm. (c) The synthesized images showing 'T', 'H', and 'U' in which each voxel is realized by tuning the wavelength ($\theta_y$) and phase ($\theta_x$).

We achieve two-dimensional beam steering by tuning the phase and wavelength. We steer the far-field beam spot from −12° to +12° along $\theta_x$ direction by tuning the phase of each channel (Fig. 4a). Across this FOV, the side lobe suppression ratio is always lower than 10 dB. By changing the output wavelength of the laser (Santec TSL 550) from 1500 nm to 1600 nm, we steer the beam spot from 5.5° to −2.5° along $\theta_y$ direction, corresponding to a tuning efficiency of 0.08 °/ nm (Fig. 4b). To demonstrate our OPA's potential in applications such as image projection, we steer the beam spot to different positions and in turn synthesize images by superimposing different frames. In this way, we obtained three images depicting 'T', 'H' and 'U' (Fig. 4c).

When we try to keep the far-field beam spot to a particular direction by maintaining the modulation voltages, the beam spot's quality will deteriorate and eventually reach an equilibrium random state. Such a phenomenon may due to the photorefractive effect. When a certain voltage is applied to a LN waveguide over the reaction time of photorefractive effect in the scale of milli-second, the carriers such as anti-site niobium defects will start to drift, leading to the redistribution of carriers over the LN waveguide. Such a redistribution of carriers will induce built-in electric fields in the LN waveguide. The superposition of these built-in electric fields and the applied electric field will induce a change in local refractive index via electro-optic effect and in turn modify the output phase of each phase modulator. The change in the output phase of each phase modulator is different because each phase modulator is under a distinct applied voltage[24, 33]. This photorefractive effect can be mitigated by annealing the LN waveguide[34] or removing the $SiO_2$ cladding layer[35]. Alternatively, the photorefractive effect can also be mitigated by steering the beam at a speed much higher than the response speed of photorefractive effect.

In summary, we demonstrated an integrated LN OPA capable of two-dimensional beam steering. The OPA exhibits a steering range of 24° and 8° in the phase and wavelength tuning directions, respectively. LN OPA's potential in achieving low power consumption, low optical loss, and high steering speed makes it an appealing candidate for many applications. The wide transparency window of LN enables LN OPA's application at other wavelengths such as visible regime for image projection.

**Funding.** This work was supported by the National Key Research and Development Program of China (2021YFA-1401000 and 2021YFB2801600), National Natural Science Foundation of China (62075114), Beijing Natural Science Foundation (4212050), and Zhuhai Industry University Research Collaboration Project (ZH-22017001210108PWC).

**Acknowledgments.** The authors thank Weiqiang Xie, Xuyue Guo for helpful discussions. The OPA chip was fabricated by Tianjin Huahuixin Technology Group Co., Ltd. Jun Mao, Tianxiang Dai, and Jianwei Wang provided the PCB and made the wire bonding connecting the OPA chip and PCB. Linjie Zhou provided the voltage source for the OPA. This work was supported by the Center of High-Performance Computing, Tsinghua University.

**Disclosures.** The authors declare no conflicts of interest

**Data availability.** Data underlying the results presented in this paper are not publicly available at this time but may be obtained from the authors upon reasonable request.